# Determination of the local concentrations of Mn interstitials and antisite defects in GaMnAs

**F. Glas, G. Patriarche, L. Largeau, and A. Lemaître**

Laboratoire de Photonique et de Nanostructures, CNRS, route de Nozay, 91460 Marcoussis, France

**Abstract:** We present a method for the determination of the local concentrations of interstitial and substitutional Mn atoms and As antisite defects in GaMnAs. The method relies on the sensitivity of the structure factors of weak reflections to the concentrations and locations of these minority constituents. High spatial resolution is obtained by combining structure factor measurement and X-ray analysis in a transmission electron microscope. We demonstrate the prevalence of interstitials with As nearest neighbors in as-grown layers.

The combination of electronic and magnetic properties in a single device opens large prospects to information technology. To this end, a particularly interesting material is GaMnAs, a GaAs crystal containing several percents of Mn atoms (for a recent review, see Ref. [1]). GaMnAs is a ferromagnetic semiconductor compatible with the well-mastered GaAs system, where Curie temperatures $T_C$ above 160 K have already been measured [2]. Boosted by predictions of possible room temperature ferromagnetism [3], a large effort involving the controlled fabrication of the material as well as the characterization and understanding of its physical properties has developed.

Basically, GaMnAs is a zinc-blende GaAs crystal where a small fraction of Mn atoms has been introduced. However, experimental and theoretical investigations have shown that this material cannot be described simply as a GaAs crystal where all Mn atoms substitute to Ga atoms to form a proper GaMnAs alloy, since additional Mn atoms can occupy interstitial sites [4,5]. Moreover, whereas GaAs grown at high temperature is highly stoechiometric, GaMnAs is grown at low temperature, a regime in which a large concentration of As atoms occupying Ga sites (antisite defects) is expected to occur. However, few structural studies so far have gone beyond the mere determination of the total Mn concentration. A notable exception is the work of Yu *et al.*, who used particle induced X-ray emission and Rutherford backscattering to determine the macroscopic concentration of Mn interstitials in GaMnAs layers [4]. These authors correlated the increase of $T_C$ upon low temperature annealing to a decrease of the concentration of these mobile atoms. Indeed, Mn interstitials and antisite defects are both believed to decrease $T_C$ by acting as donors compensating the holes provided by the substitutional Mn atoms, which induce ferromagnetism [1,4,5]. Since the electronic, magnetic and structural [5,6] properties of GaMnAs appear highly sensitive to these species, it is desirable to quantify them and, if possible, locally. In this letter, we develop a simple method for measuring jointly and with high spatial resolution the concentrations of all relevant species, namely substitutional and interstitial Mn and antisite defects. The body of the letter describes our method, which we illustrate by analyzing a particular area via Table 1 and the figures.

To describe the GaMnAs crystal, we take the GaAs matrix as a reference. We choose a face centered cubic (fcc) unit cell with an As atom at $(0,0,0)$ and a Ga atom at $\mathbf{R} = a\,(¼,¼,¼)$, where $a$ is the lattice parameter of the alloy (the epitaxial strain will be discussed later). Whereas in GaAs, the As and Ga atoms wholly occupy these two 'matrix' fcc sublattices, in GaMnAs the Mn atoms may occupy either Ga sites or interstitial sites.

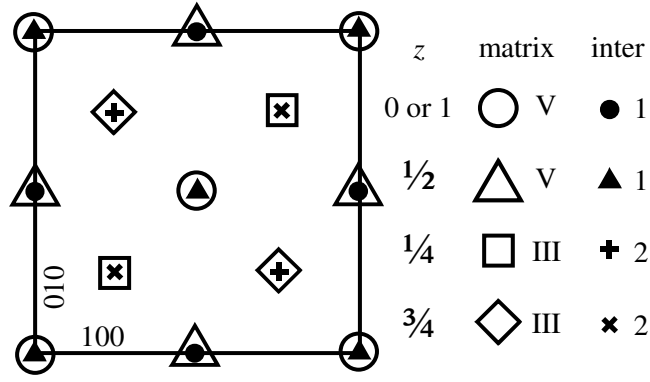

FIG 1: Sites in the fcc unit cell. III, V, 1 and 2 indicate respectively Ga and As matrix sites and type-1 and type-2 interstitial (inter) sites. $z$ is the reduced coordinate along the perpendicular <100> direction.

Moreover, there are two types of interstitial sites located on two 'interstitial' fcc lattices translated from the As sublattice by $2\mathbf{R}$ and $3\mathbf{R}$ (Fig. 1), which are the networks of tetrahedral interstitials sites of the Ga and As fcc sublattices, respectively. A Mn atom has four Ga nearest neighbors (NNs) in the former (type 1) and four As NNs in the latter (type 2).

We lack detailed estimations of the abundances of these various defects. Experimentally, Yu *et al.* measured the total concentration of interstitials without however distinguishing the two types of sites [4]. Scanning tunneling microscopy might seem able to provide appropriate sensitivity and resolution. However, data interpretation is not straightforward: some studies identify only antisite defects [7,8] whereas others also detect interstitials [9,10], and no concentration measurement has been reported. Conversely, most theoretical studies (*e.g.* Refs. [5] and [11]) assume that only type 2 interstitials exist, although Edmonds *et al.* have recently considered the two types [12].

To solve this problem, we first fabricated 300 nm thick ferromagnetic GaMnAs layers by molecular beam epitaxy on GaAs (001) substrates. Each GaMnAs layer was grown at 270°C after depositing a GaAs buffer layer at 600°C. Unannealed samples were then studied by transmission electron microscopy (TEM) using 200 keV electrons.

Figure 2 is a typical TEM dark field (DF) image formed by selecting the 002 diffracted beam. It readily appears that the 002 intensity is much lower for the GaMnAs layer than for the GaAs buffer layer or substrate. We analyze such images by using the two-beam kinematical approximation, whereby the intensity at exact Bragg incidence for reflection $\mathbf{g}$ is proportional to $t^2 V_c^{-2} |F_{\mathbf{g}}|^2$, with $t$ the specimen thickness, $V_c$ the unit cell volume and $F_{\mathbf{g}}$ the structure factor (SF) of reflection $\mathbf{g}$ [13]. Hence, if $t$ is uniform, the image maps the SF variations, with a small correction due to possible variations of $V_c$. In practice, we measure the ratio $\rho_{\mathbf{g}}$ of the DF intensities recorded locally in GaMnAs and in neighboring GaAs (Table 1). The kinematic approximation is justified by the low values of the SFs ($F_{002} \approx 0.2633$ nm for GaAs), corresponding to extinction distances [13] $\xi_{\mathbf{g}} \gg t$ ($\xi_{002} \approx 865$ nm in GaAs), and by calculations showing that for $\mathbf{g} = 002$ dynamical effects are very small in the closely related InGaAs alloys [14].

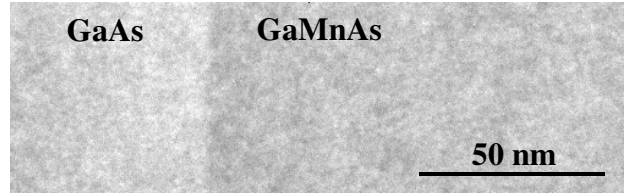

FIG 2: TEM 002 DF image of a GaMnAs layer grown on a GaAs substrate.

For quantitative analysis, we consider six atomic species: Ga and As at their proper sites, As antisites, substitutional Mn and the two types of interstitial Mn. Our TEM studies rule out the presence of MnAs precipitates (surmised by Yu et al. [4] in annealed samples) and we ignore possible As vacancies, which have been detected only in non-ferromagnetic samples [15]. Hence, assuming random occupation of the relevant sites, neglecting possible static atomic displacements [16] and leaving aside the Debye-Waller factors (which will almost cancel each other when we compute ratios of DF intensities), the SF for reflection $\mathbf{g}$ is:

$$F_{\mathbf{g}} = 4\{c_{As}^V f_{As} + e^{i\varphi}[c_{Ga}^{III} f_{Ga} + c_{As}^{III} f_{As} + c_{Mn}^s f_{Mn}] + e^{2i\varphi} c_{Mn}^1 f_{Mn} + e^{3i\varphi} c_{Mn}^2 f_{Mn}\} \quad (1),$$

where $f_A$ is the appropriate atomic scattering amplitude (ASA) for atom $A$ [17], $c_A^\sigma$ is the concentration of matrix atom $A$ (Ga or As) on matrix sublattice $\sigma$ (III or V), $c_{Mn}^s$, $c_{Mn}^1$ and $c_{Mn}^2$ are the concentrations of Mn atoms occupying respectively the subsitutional and the two types of interstitial sites, and $\varphi = 2\pi \mathbf{g}.\mathbf{R}$. Here, a unit concentration corresponds to the occupation of all the sites of one of the fcc matrix sublattices. Hence, the equations describing the occupations of these two sublattices are $c_{As}^V = 1$ and $c_{Ga}^{III} + c_{As}^{III} + c_{Mn}^s = 1$. Except $c_{As}^V$ and $c_{Ga}^{III}$, which are close to 1, all concentrations are at most a few percents.

As in GaAs and III-V alloys, two types of reflections exist [16]. For the 'strong' reflections, $\varphi = 0$ or $\pm \pi/2$, and the relative variations of the SF depend little on the concentrations of the minority atomic species. Instead, we use 'weak' reflections (*e.g.* 002), whose indices verify $h + k + l = 4n + 2$ with *n* integer, for which $\varphi = \pi$ and (1) rewrites:

$$F_{\mathbf{g}} = 4\left\{ c_{As}^V f_{As} - \left[ c_{Ga}^{III} f_{Ga} + c_{As}^{III} f_{As} + c_{Mn}^s f_{Mn} \right] + c_{Mn}^1 f_{Mn} - c_{Mn}^2 f_{Mn} \right\} \qquad (2).$$

For a weak reflection, in GaAs, $F_{\mathbf{g}} = 4(f_{As} - f_{Ga})$ is low because the contributions of Ga and As (which have close atomic numbers) nearly cancel each other. This remains true in GaMnAs for the contributions of the atoms belonging to the two matrix sublattices, since Mn is also close to Ga and As in atomic number. However, there is a further contribution from the interstitials. Numerically, we find:

$$F_{002}(\text{GaMnAs}) = F_{002}(\text{GaAs}) + 4\left[ 0.003 c_{Mn}^s - 0.066 c_{As}^{III} + 0.551 \left( c_{Mn}^1 - c_{Mn}^2 \right) \right] \quad (\text{nm}) \qquad (3),$$

so that $F_{002}$ is virtually independent of the substitutional Mn concentration and 8 times more sensitive to the difference of interstitial Mn concentrations than to the concentration of As antisites. To calculate the ASAs yielding the numerical coefficients of Eq. (2), we assumed equal 002 scattering angles for GaMnAs and bulk GaAs; however, correcting this angle for tetragonal strain (see below) makes little difference.

If, as a first approximation, we neglect the contributions of the substitutional Mn and As antisites to the SF, Eq. (3) becomes $F_{002}(\text{GaMnAs})/F_{002}(\text{GaAs}) \approx 1 + 8.38 \left( c_{Mn}^1 - c_{Mn}^2 \right)$. Hence, $F_{002}$ and the 002 DF intensity are highly sensitive to the interstitials. Moreover, the two types of interstitials have opposite effects on the SF: type 1 interstitials increase it, whereas type 2 interstitials reduce it. Experimentally, we find that the 002 DF intensity is systematically lower in GaMnAs than in GaAs (Fig. 2 and Table 1): this readily demonstrates that type 2 interstitials are the most common interstitials. Our technique is related to, but different from, the use of weak reflections to measure concentrations in stoechiometric ternary III-V alloys [16,18], which is restricted to matrix atoms. However, both techniques have the same contrast detection limits, which we estimate to about 1% in terms of the ratio $\rho_{002}$, limited mainly by the noise of the recording system. Using the previous approximation, this gives an interstitial detection limit of only $6 \times 10^{-4}$ for $\left| c_{Mn}^1 - c_{Mn}^2 \right|$.

**Table 1**: Ratio $\rho_{002}$ of the 002 DF intensities for GaMnAs and GaAs and ratios $r_G$ and $r_M$ of the Ga and Mn concentrations to that of As, measured in the TEM; lattice parameter $a_\perp$ along [001] measured by XRD; difference $(\tilde{c}_{Mn}^1 - \tilde{c}_{Mn}^2)$ of the concentrations of the two types of Mn interstitials estimated by neglecting the other minority species; concentrations $c_{Mn}^s$ and $c_{Mn}^2$ (substitutional and interstitial Mn) and $c_{As}^{III}$ (antisite defects) calculated by assuming no type 1 interstitial. Except $a_\perp$, all values pertain to a given analyzed area.

| $\rho_{002}$ | $r_G$ | $r_M$ | $a_\perp$ (nm) | $(\tilde{c}_{Mn}^1 - \tilde{c}_{Mn}^2)$ (%) | $c_{Mn}^s$ (%) | $c_{Mn}^2$ (%) | $c_{As}^{III}$ (%) |
|---|---|---|---|---|---|---|---|
| 0.81 | 0.877 | 0.057 | 0.56814 | 1.14 | 5.18 | 0.76 | 3.77 |
| ± 0.01 | ± 0.014 | ± 0.0026 | | | ± 0.33 | ± 0.013 | ± 0.04 |

These are remarkable results for a TEM method. Indeed, TEM is usually unable to detect (not to speak about measuring) a concentration of minor non-matrix constituent. Moreover, site-sensitivity is uncommon and requires specific TEM techniques involving the setting up of precisely controlled channeling conditions [19]. Here, we show that in favorable cases the standard DF technique may yield such information and sensitivity.

The bold approximation made above ($c_{Mn}^s = 0$, $c_{As}^{III} = 0$) enables us to calculate from the measured intensity ratio a first estimate $\tilde{c}_{Mn}^1 - \tilde{c}_{Mn}^2$ of the difference of interstitial concentrations (Table 1). However, it is desirable to characterize the material more fully. Indeed, Yu *et al.* [4] found a total interstitial concentration of about 1/6 of the total Mn concentration $c_{Mn}$ and Mašek *et al.* [5] estimated theoretically an antisite concentration of about $c_{Mn}/2$, so that in Eq. (3) the contribution of the antisites to $F_{002}$ might not be negligible. Fortunately, it is possible fully to analyze the material by using additional data gathered from the same specimen area in the same instrument. To this end, we perform TEM energy dispersive X-ray (EDX) microanalysis. Using standard correction procedures, these measurements provide the ratios $r_G$ and $r_M$ of the atomic concentrations of Ga and Mn to that of As (Table 1). This yields two more relations between the concentrations, namely $c_{Ga}^{III} = r_G(1 + c_{As}^{III})$ and $(c_{Mn}^s + c_{Mn}^1 + c_{Mn}^2) = r_M(1 + c_{As}^{III})$. We now have five relations (the latter two and Eq. (2), constituting the experimental results, and the two matrix sublattice occupation equations) between six unknowns, from which we deduce a single relation (valid for any weak reflection) between two independent variables, *e.g.* $c_{Mn}$ and $c_{Mn}^2$:

$$F_{\mathbf{g}} = 4\left\{ 2\,(f_{As} - 2f_{Mn}) - \left[\frac{(f_{As} - 2f_{Mn}) + r_G\,(f_{Ga} - 2f_{Mn})}{r_M} + \left(2\frac{c_{Mn}^2}{c_{Mn}} - 1\right)f_{Mn}\right]c_{Mn}\right\} \quad (4)$$

All other concentrations are easily calculated from these two variables: $c_{Ga}^{III} = r_G\,c_{Mn}/r_M$, $c_{As}^{III} = c_{Mn}/r_M - 1$, $c_{Mn}^s = 2 - (r_G + 1)c_{Mn}/r_M$ and $c_{Mn}^1 = c_{Mn} - c_{Mn}^s - c_{Mn}^2$. In addition, X-ray diffraction (XRD) yields the lattice parameter $a_\perp$ along the growth direction, from which the actual 002 scattering angle and then the ASAs can be accurately evaluated. Moreover, TEM indicates that the layer is coherently grown on GaAs, so that $V_c = a_0^2\,a_\perp$, where $a_0 = 0.565325$ nm is the lattice parameter of GaAs.

Figure 3 shows the variations of $|F_{002}(\text{GaMnAs})/F_{002}(\text{GaAs})|^2$ with $c_{Mn}$ and the fraction $c_{Mn}^2/c_{Mn}$, calculated from Eq. (4) by using the EDX results (Table 1), in the ranges compatible with the latter. From the previous discussion, this quantity is equal to the experimentally determined quantity $(a_\perp/a_0)^2\,\rho_{002}$. The dashed line corresponds to the area analyzed (Table 1). With no further hypothesis, each concentration $c_{Mn}$ corresponds to a unique given set of concentrations of the minority species (Fig. 4). In any case, as expected from the discussion of the contrast using Eq. (3), the interstitials are mainly of type 2.

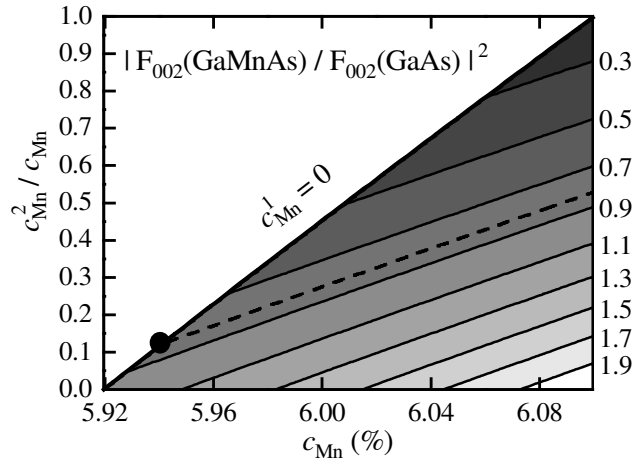

FIG 3: Map of the variations of the ratio $|F_{002}(\text{GaMnAs})/F_{002}(\text{GaAs})|^2$ with $c_{Mn}$ and $c_{Mn}^2/c_{Mn}$, calculated by using the EDX results of Table 1. Lines of equal ratio for the values given in the right margin (full curves) and for the experimental value (Table 1; dashed curve). Thick diagonal line: locus $c_{Mn}^1 = 0$. Dot: experimental result assuming $c_{Mn}^1 = 0$.

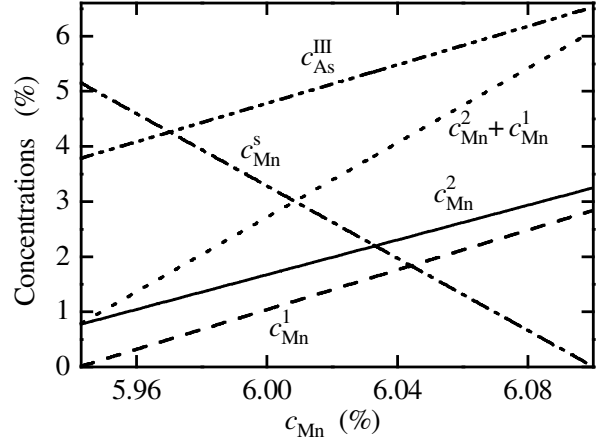

FIG 4: Variations with the total Mn concentration of the concentrations of Mn interstitial and substitutional atoms and of antisite defects, compatible with the TEM results of Table 1.

All data in Fig. 4 are compatible with our experiments and yield remarkably narrow ranges for the total Mn concentration and, as expected from our first approximation, for the difference of the concentrations of the two types of interstitials. However, from previous studies, high fractions of interstitials and low fractions of substitutional Mn (found in most of Fig. 4, apart from its left section) appear unlikely [4,5]. To proceed further, we thus assume that only type 2 interstitials exist ($c_{Mn}^1 = 0$, upper border of the domain shown in Fig. 3). This hypothesis, already adopted by Mašek *et al.* [5], is substantiated by Edmonds *et al.*, who calculated that a Mn atom has a lower energy in type 2 sites than in type 1 sites, unless it pairs with a substitutional Mn atom [12]. Then, $c_{Mn} = r_M \left(c_{Mn}^2 + 2\right)/(1 + r_G + r_M)$ and:

$$c_{Mn}^2 = \frac{2 r_G (f_{As} - f_{Ga}) + 2 r_M (f_{As} - f_{Mn}) - \tfrac{1}{4} (1 + r_G + r_M) F_\mathbf{g}}{f_{As} + r_G f_{Ga} + r_M f_{Mn}} .$$

The determination of all relevant concentrations follows (Table 1 and dot in Fig. 3). The local fraction of type 2 interstitial Mn (13% of the total Mn concentration) is only slightly smaller than the global fraction of interstitials (types 1 and 2) measured macroscopically by Yu *et al.* in as-grown samples (17%) [4]. Using the elastic constants of GaAs, we deduce from $a_\perp$ (Table 1) a bulk lattice parameter $a = 0.5668$ nm whereas, for the concentrations of Table 1, Eq. (5) of Ref. [5] yields $a = 0.5683$ nm. On the other hand, the concentration of antisites is close to $c_{Mn}/2$, in agreement with the expectations of Mašek *et al.* [5]. We however measured smaller concentrations of antisites in other samples, and the values calculated from Eq. (5) of Ref. [5] are then in excellent agreement with the measured lattice parameter. This is to our knowledge the first validation of the latter equation taking into account all minority species (Kuryliszyn-Kudelska *et al.* discussed its validity only as regards the change of lattice parameter upon estimated changes of interstitial concentrations induced by annealing [6]).

SF measurement might also be performed accurately by XRD. This technique would be similarly sensitive to the concentrations and types of interstitials, since the reflections which have low SF for electrons also have low SF for X-rays [16] and the phases of the atomic contributions to the SF are the same. However, the present TEM method is unique, not only in offering more information (namely multiple concentration measurement at a given point), but also because it can achieve high spatial resolution. Indeed, DF imaging has an aperture-limited resolution of about 0.5 nm, whereas the resolution of the EDX analysis, limited by probe size and beam spreading, may be nanometric in very thin areas. However, better statistics are obtained by analyzing larger and thicker areas, which is possible here since the TEM contrast of our epitaxial layers is usually very uniform, although thin sublayers with markedly different contrast, indicating a different concentration of interstitials, appear locally.

In summary, we have demonstrated that, by combining imaging and analytical TEM techniques, it is possible to measure with high spatial resolution the local concentrations of substitutional and interstitial Mn atoms and of As antisite defects in GaMnAs. These determinations rely on a detailed analysis of the variations of the 002 structure factor with the concentrations of these minority species. The structure factor is highly sensitive to the interstitials but only weakly to substitutional Mn and antisite defects. Moreover, the two types of interstitials make it vary in opposite directions. In the unannealed sample examined, the interstitials with As nearest neighbors dominate. Analyses of as-grown and annealed layers will be reported elsewhere.

This work was supported by Région Ile de France, SESAME project No 1377 and Conseil Général de l'Essonne.